\documentclass[aps,secnumarabic,showpacs,nobibnotes,showkeys,superscriptaddress,
amssymb,epsfigs,amsfonts,nofootinbib,amsmath,apsfonts,eqsecnum]{revtex4}
\usepackage{graphicx}
\usepackage{dcolumn}
\usepackage{bm}
\begin{document}
\title{\Large\bf Calculation of the Masses and the Running Masses of the
 Quarks and Leptons from Electroweak to Supersymmetric Grand Unification Mass} 
\author{B. B. Deo} 
\email{bdeo@iopb.res.in}
\affiliation{Department of Physics , Utkal University, Bhubaneswar-751004, India.}
\author{P. K. Mishra}
\affiliation{Department of Physics, S.C.S. College, Puri-752001,  India}
\date{\today}
\begin{abstract}
We make a systematic theoretical analysis of the masses of the fermions 
and their variation 
with energy by solving the one loop renormalization group equation (RGE) 
in the Minimal Supersymmetric Standard Model (MSSM). A deceptibly simple 
common mass for all the fermions 
around 115 GeV at GUT scale has been found by Deo and Maharana. Here we 
undertake the unfinished but the important task of calculating the electroweak  
masses of the fermions at different energies. The proposed parametric 
unification mass and group theoretic constants 
for the model are known. The mass of the top quark and its descent is studied 
by an approximate method very carefully. 
We find that the Ramond, Roberts and Ross value of the Wolfenstein parameter 
is reproduced and is nearly equal to 0.22. When raised to integral powers and
multiplied by 115 GeV, the whole mass 
spectra of the remaining eleven  fermions are obtained within 
experimental errors. We deduce formulae for the masses
and plot them for all the 12 fermions from 
$t=log \frac{\mu}{M_z}=0$ to $t_X=33$; the GUT mass being 
$M_X=2.2\times 10^{16}$ GeV.
\end{abstract}
\maketitle

\section{Introduction}
Pati and Salam\cite{a1} pioneered the idea that leptons are the fourth colour;
quarks and leptons should be brought under the same umbrella of one group so
that all forces ( except gravity) can be understood in terms of one unifying 
force parameter near the Planck scale. The six leptons should be treated at par
with the six coloured quarks. The elementary constituents of matter would become
twelve only. With the availability of enormous data from the high
energy accelerators, phenomenological analysis backed by imaginative theories, it
has been found that the  familiar standard model, which is the product group
$SU_{C}(3)\otimes SU_{L}(2)\otimes U_{Y}(1)$, can explain all of them successfully.
This standard model is characterised by three coupling strengths of  the weak, the
electromagnetic and the strong interactions. It was conceived that all the 
three couplings should run, i.e., they change with energy, and eventually become
one at the grand unified scale. This could be made possible by solving the relevant 
RGE. The coefficients of the theory are specified by the group structure alone.
Many theoretical models were investigated and only a few years back, it has become
essential to bring in supersymmetry. In the minimal version of the MSSM, the beta 
coefficients are such that they run the coupling 
strengths to a single value $\frac{4\pi}{g_{U}^{2}}=\alpha_{GUT}\simeq \frac{1}{25}$ 
at a mass of $M_{X}=2.2\times 10^{16} $ GeV. This has been the most attractive
result in the current investigations in gauge theories\cite{a2}.

Next in order, the major challenge in particle physics, was and is the theoretical
derivation of the mass spectrum of the quarks and leptons in the same successful
MSSM theory. In this model, all the masses of the fermions and the mixing angles were 
being chosen arbitrarily to account for  the 19 free parameters of the theory. 
In the absence of such a fundamental theory, it has been in vogue to pursue the method 
which has been known as `textures analysis'. After finding a suitable texture, 
one can investigate further and possibly obtain  unification mass parameters 
for all the fermions as a generalisation of the hypothesis by 
Georgi et al\cite{a3}.
Eventually, one can predict their individual masses by using the analytical MSSM
group coefficients for a RGE for the masses of the fermions. As yet, this has not 
been successful. 

One of the 13 parameters of the standard model was first predicted in 1974 by
Gaillard and Lee\cite{a4}. Then came the work of the popular mass matrix ansatz
of Fritzsch\cite{a5}. A complete listing of textures and their relevance to
experimental findings were made by Ramond, Roberts and Ross ( RRR)\cite{a6}.
Since they also took the help of RGE, we shall present the essential results
of their work in this section. Before that, we present the MSSM Lagrangian
\begin{equation}
{\cal{L}}_M=\bar{Q}_LM_U\Phi_uU_R+\bar{Q}_LM_D\Phi_dD_R+\bar{l}_LM_E\Phi_dE_R+
\bar{\nu}_LM_N\Phi_u\nu_R+h.c.
\end{equation}
We ignore the sparticles and consider the rigid part of the Lagrangian given by
Demir~\cite{Demir04}.

In the renormalisation schemes, the Yukawa couplings $M_F(t)$ and v.e.v. $v_F(t)$
change with energy separately. The Dirac masses are given by
\begin{equation}
m_F(t)=v_F(t)M_F(t)
\end{equation}
The masses considered above are not the masses of the flavor eigenstates
of the model.

The one loop RGE, as written by Grzadkowski, Lindner and Theisen \cite{a9} 
for $M_Y(t)$ are
\begin{equation}
16\pi^2\frac{dM_Y}{dt~~}=\left (-G_Y+T_Y+S_Y\right ) M_Y(t),\label{e3a}
\end{equation}
$t=log(\mu/M_Z)$, $\mu$ is the renormalisation point and  $M_Z$ is the mass 
of the Z-boson. Here Y=U stands for U-quarks (F=1,2,3): up(u), charm(c) and top(t); Y=D stands
for Down quarks (F=4,5,6): bottom(b), strange(s) and down(d); Y=N stands for Neutrinos
(F=7,8,9):~$\nu_e,~\nu_{\mu}$, and $\nu_{\tau}$, and Y=E stands for electrons(F=10,11,12): 
~e,~$\mu$ and $\tau$. $G_Y(t)$ contains the gauge coupling terms, given in Section-2 and
\begin{equation}
T_U= Tr\left ( 3M_UM^{\dagger}_U +M_NM^{\dagger}_N\right ),~~~~~T_D=
Tr\left ( 3M_DM^{\dagger}_D +M_EM^{\dagger}_E\right ),~~T_U=T_N, T_E=T_D,
\end{equation}
and
\begin{equation}
S_U=3M_UM^{\dagger}_U +M_DM^{\dagger}_D,~~S_D=3M_DM^{\dagger}_D +M_UM^{\dagger}_U,~~
S_E=3M_EM^{\dagger}_E +M_NM^{\dagger}_N,~~S_N=3M_NM^{\dagger}_N 
+M_EM^{\dagger}_E.\label{e3c}
\end{equation}
To find the couplings, we have to solve the twelve differential equations and  
determine all the couplings/masses from only one value of coupling at $t=t_X$ or 
$ M_F(M_X)$. First, we turn our attention to the mass of the top quark.
Incidentally, Pendleton and Ross~\cite{pendleton}, Faraggi\cite{faraggi}, 
and some others have predicted the value of the mass of the top quark which 
was around 175 GeV, even before the top was discovered.

In this paper,
all the one loop equations are solved following the same method used for gauge
coupling RGE. A particular $M_F(M_Z)$ is obtained relating to 
$M_F(M_X)$ and other fermions. Considering the top, first we make 
'heavy top integral' approximation stated below. Eventhough we
aim at single input value $M_F(M_X)=M_U$ , which is independent of F, the 
integrals of the solutions for the masses are such that
\begin{equation}
\int M^2_{top}(\tau)d\tau \gg \int M^2_{Q\neq top}(\tau)d\tau \ggg 
\int M^2_{lepton}(\tau)d\tau\label{e6}
\end{equation}
Neglect of the integrals other than the top quark, gives a very good result for the
$M_{top}(M_X) \simeq M_U$. Following this approximation procedure, we find from other
eleven equations 
that $M_{top}(M_X)\simeq M_F(M_X) \simeq M_U \simeq$ 115 GeV for 
all the fermions. 

We choose $ M_U \simeq$ 115 GeV as the only input. Furthermore, MSSM has two Higgs.
\begin{eqnarray}
<\phi_U> = v_u(t)&=&v(t) Sin\beta(t)\\
<\phi_D> = v_d(t)&=&v(t) Cos\beta(t)\\
v^2(t)&=& v^2_d(t)+v^2_u(t)\\
tan\beta(t)&=&\frac{v_u(t)}{v_d(t)}.
\end{eqnarray}
It is to be noted that $v(t)$ is the vacuum expectation value of single Higgs
of the Standard Model. $v(M_Z)$=246 GeV, $Sin\beta_{SM}(M_Z)=\frac{1}{\sqrt{2}}$.
We note that $v_0=\frac{246}{\sqrt{2}}$ GeV=174 GeV, which is close to 
$M_{top}(M_Z)\cong 175$ GeV. For simplicity, we shall use this as the unit of energy
whenever unspecified.

The one loop RGE are 
\begin{eqnarray}
16\pi^2 \frac{dv_u}{dt}&=&
\left( \frac{3}{20}g_1^2+ \frac{3}{4}g_2^2-Tr(3M_UM_U^{\dagger})\right)v_u,\label{e10}\\
16\pi^2 \frac{dv_d}{dt}&=&
\left( \frac{3}{20}g_1^2+ \frac{3}{4}g_2^2-Tr(3M_DM_D^{\dagger})\right)v_d.\label{e11}
\end{eqnarray}
The calculation of 
\begin{equation}
tan\beta(t)=\frac{v_u(t)}{v_d(t)},\label{e12}
\end{equation}
in MSSM has attracted considerable attention, there exist extensive literature,
most of them are given in reference\cite{parida}. We note that below the mass scale 
$M_Z$ of the normal SM, one Higgs v.e.v. satisfies the one loop equation
\begin{equation}
16\pi^2 \frac{dv_{\scriptsize{SM}}}{dt}=
\left( \frac{3}{20}g_1^2+ \frac{3}{4}g_2^2-Tr(3M_UM_U^{\dagger}
+3M_DM_D^{\dagger})\right)v_{\small SM}.
\end{equation}
To maintain continuity, we assume that at $t$=0,
\begin{equation}
tan\beta(M_Z)=tan\beta_{SM}(M_Z)=1.
\end{equation}
Using the approximation given in equation (\ref{e6}), we get from equations (\ref{e10}),
(\ref{e11}) and (\ref{e12}),
\begin{equation}
tan\beta(t)\cong exp\left ( -\frac{3}{16\pi^2}\int_0^t 
\left [M^2_{top}(\tau)-M^2_{bottom}(\tau)\right]d\tau\right ).
\end{equation}
Using the results given in section-3, equation(\ref{e75}), 
we find that $tan\beta(t)$ is a slowly varying
function of $t$ and drops from 1 at 91 GeV to 0.9 at $10^{16}$ GeV. So the calculation 
is much simplified if we take $Sin\beta=\frac{1}{\sqrt{2}}=Cos\beta$. This is
not ruled out by experiment\cite{a9}.

The authors (RRR), noted that there are 
only six possible forms of symmetric mass matrices with three non-zero eigenvalues 
and three texture zeroes, capable of describing the hierarchy of up or down quark 
mass matrices. Those are
\begin{eqnarray}
T_{1}=
\left (
\begin{array}{ccc}
 a_1 & 0 & 0\\
0 & b_1 & 0\\
0 & 0 & c_1
\end{array}
\right ),~~~~~
T_2=
\left (
\begin{array}{ccc}
 0 & a_2 & 0\\
a_2 & b_2 & 0\\
0 & 0 & c_2
\end{array}
\right ),~~~~~~~~~~~
T_3=
\left (
\begin{array}{ccc}
 a_3 & 0 & 0\\
0 & 0 & b_3\\
0 & b_3 & c_3
\end{array}
\right ),\nonumber\\
T_{4}=
\left (
\begin{array}{ccc}
 0 & 0 & a_4\\
0 & b_4 & 0\\
a_4 & 0 & c_4
\end{array}
\right ),~~~~~
T_{5}=
\left (
\begin{array}{ccc}
 0 & a_5 & 0\\
a_5 & 0 & b_5\\
0 & b_5 & c_5
\end{array}
\right )~~~~\texttt{and}~~~
T_{6}=
\left (
\begin{array}{ccc}
 0 & a_6 & b_6\\
a_6 & 0 & 0\\
b_6 & 0 & c_6
\end{array}
\right ).
\end{eqnarray}
$T_5$ was the one first pioneered by Fritzsch\cite{a5}. More recently Dimopoulos,
Hall and Raby\cite{a7} have analysed and included the leptons following
Georgi\cite{a3} and solved the MSSM RGE with some degree of success. RRR tried 
to analyse all the cases and put them in a CKM matrix form, proposed by 
Wolfenstein\cite{a8} and diagonalise them to the texture types.
\begin{eqnarray}
V_{CKM}&=&
\left (
\begin{array}{ccc}
c_1c_2-s_1s_2e^{-i\phi}& s_1+c_1s_2e^{-i\phi}                  & s_2(s_3-s_4)\\
-c_1s_2-s_1e^{-i\phi}  &-s_1s_2+(c_1c_2c_3c_4+s_1s_4)e^{-i\phi}&s_3-s_4\\
s_1(s_3-s_4)           &-c_1(s_3-s_4)                          & (c_3c_4+s_1s_4)e^{i\phi}\\
\end{array}\label{eq69a}
\right )\\
&=&
\left (
\begin{array}{ccc}
 \lambda^{2}/2             & \lambda         & A\lambda^{3}(\rho + i\eta)\\
-\lambda                   & 1-\lambda^{2}/2 & A\lambda^2\\
A\lambda^{3}(1-\rho+i\eta) & -A\lambda^2     & 1
\end{array}
\right ).\label{eq2}
\end{eqnarray}
Here $s_i,~c_i$ (i=1,..,4)are the sines and cosines of mixing angles.
From the identities given by Dimopoulos, Hall and Raby\cite{a7}, following from
equations (\ref{eq69a}) and (\ref{eq2})
\begin{equation}
\lambda=(s_1^2+s_2^2+s_1s_2~Cos\phi)^{1/2}.
\end{equation}
$\phi$ is the CKM phase angle
and
\begin{equation}
s_1=\left (\frac{M_d}{M_s}\right )^{\frac{1}{2}}=\lambda,~~~
s_2=\left (\frac{M_u}{M_c}\right )^{\frac{1}{2}}=\lambda^2;~~~~
s_4=\left (\frac{M_dM_s}{M_b^2}\right )^{\frac{1}{2}}=\lambda^3.
\end{equation}
As will be discussed later, the $Cos\phi$ defined in ~\cite{a7}, we shall get
\begin{equation}
Cos\phi=-\lambda/2,
\end{equation}
and                                                                               
\begin{equation}
s_3-s_4=\lambda^2A(t)
\end{equation}
where A(t) depends on $t$. 
                                                                                             
The small expansion parameter is $\lambda \simeq~~$0.2 and $A\simeq 0.9\pm0.1$.
Olechowski and Poroski\cite{a9} were the first to write down the RG Equations for the
parameters
\begin{eqnarray}
16\pi^2\frac{d|J_{c\rho}|}{dt}&=&-~3~c~(h_t^2~+~h_b^2)|J_{c\rho}|,\label{eq4b}\\
16\pi^2\frac{dA}{dt}&=&-\frac{3}{2}~c~(h_t^2~+~h_b^2)A,\label{eq4}\\
\frac{d\lambda}{dt}&=&0,\label{eq4a}\\
\frac{d\rho}{dt}&=&0,\label{eq4c}\\
\frac{d\eta}{dt}&=&0.\label{eq4d}
\end{eqnarray}
Here $c~=~2/3$ for MSSM and $J_{c\rho}$ is the irreducible phase of the
CKM matrix. To arrive at equation (\ref{eq4}), we  equate
\begin{equation}
A(t)=\left [ \frac{M_{bottom}(t)M_{top}(t)}{M_{top}^2(M_X)}\right ]^{-1/7}\label{eq85a}.
\end{equation}
Using one loop RG equations (\ref{e3a}) to (\ref{e3c}), we get
\begin{equation}
A(M_X)=1.00~~~~\text{and}~~~ A(M_Z)=1.474.
\end{equation}
RRR have made a complete listing and analysis. They arrived at the value
$\lambda=$=0.22 and the oft quoted result,
\begin{eqnarray}
m_{\tau}~:~m_{\mu}~:~m_{e}~~=~~m_b~:~m_s~:~m_d~~=~~1~:~\lambda^2~:~\lambda^4 \nonumber\\
m_t~:~m_c~:~m_u~~=~~1~:~\lambda^4~:~\lambda^8,\label{eq8}
\end{eqnarray}
which is very well satisfied by experimentally found masses. This has been
thought to be  like a miracle.

We shall take the masses of the twelve fermions  to be real, wherever necessary. 
In section-2, 
we shall write  the RG Equation with the MSSM coefficients and give the one loop exact
solutions. In section-3, the unification mass for all fermions at GUT scale ranging
from 113 GeV to 125 GeV as computed by Deo and Maharana\cite{a10}, will be discussed
as a follow up of their letter. An expression for Wolfenstein parameter
in terms of RGE coefficients is given in section-4. The reason to raise
this parameter by integers $n_F$ to obtain fermion masses except the top,
$M_F$ equaling $\lambda^{n_F}$ multiplied by the parametric unification mass 115 GeV, 
is given. In sections-5,6 and 7, we suggest an alternative method of 
calculation of experimental
masses of all fermions by a suitable self contained procedure which
includes the gauge couplings as well. The equation for the running of all
the masses of all the fermions is given in section-8. The results are given in 
tables and are also shown graphically for greater clarity. 
The concluding remarks are given in the section-9.

\section{Renormalisation Group Equations and Solutions}
As stated, the gauge sector of the standard model is characterised by 
three coupling constants $g_3$, $g_2$ and $g_1$ of $SU_C(3)\otimes SU_L(2)
\otimes U_Y(1)$, respectively. However, these couplings are not constants, they
change with energy/mass values. The nature of variation is given by the solutions of the
RG Equations. The coefficients are calculated by the specific nature of the 
Standard Model group. For MSSM, the three couplings at mass 
$M_X=2.2\times 10^{16}$ GeV\cite{a11}
unite to a unified coupling constant $g_U^2/4\pi$=1/24.6. Here, the supersymmetry
descends from $M_X$ down to $M_Z\simeq$~~91 GeV, as suggested by Witten.
 The RG Equations for the couplings in the lowest order are given by
\begin{equation}
16\pi^2\frac{dg_i(t)}{dt}~=~c_ig_i^3(t),~~~~ i=1,2,3.\label{eq9}
\end{equation}
The coefficients are $c_1$~=~6.6, $c_2$~=1,~~$ c_3$~=~-3. The first two
coefficients are positive, indicating that $U_Y(1)$ and $SU_L(2)$ are
 not asymptotically free, whereas the $SU_C(3)$ colour group is free
and makes the entire product group asymptotically free . This implies
that in a perturbative formulation, the higher order contributions are
small and can be neglected. Therefore, we shall use equation(\ref{eq9}) 
only in the gauge sector, with two Higgs\cite{a11}.

The running parameter $t$ is defined as $t= log_e \frac{\mu}{M_Z}$ so
that it varies from 0 to $log_e \frac{M_X}{M_Z}\simeq 33$. The solution to
RG Equation(\ref{eq9}) is 
\begin{equation}
\frac{4\pi}{g_i^2(t)}~=~\frac{4\pi}{g_i^2}-\frac{c_i}{2\pi}t\label{eq10}.
\end{equation}
Here, $g_i^2=g_i^2(0)$ are the coupling strengths in the electroweak scale
$M_Z$. Taking the value of $M_X=2.2\times 10^{16}$ GeV and $\frac{4\pi}{g_U^2}
=24.6$, we calculate the values of $\frac{4\pi}{g_1^2}=59.24$,
$\frac{4\pi}{g_2^2}=29.85$ and $\frac{4\pi}{g_3^2}=8.85$. These are 
consistent with the experimental results. Thus the three coupling
strengths are descendants of one coupling constant $g_U$.

Taking the clue from equation(\ref{eq9}), we rewrite the Yukawa sector SUSY RG
Equations given earlier in equation (\ref{e3a}), which had also been written 
by Babu\cite{a12} following Georgi and Glashow, and Eitchen et al\cite{a13} 
in the following way;
\begin{eqnarray}
16\pi^2\frac{dM_F(t)}{dt}&=&A_FM_F^3(t)+[Y_F(t)-G_F(t)]M_F(t)\label{eq11}\\
&=&A_FM_F^3(t)+Z_F(t)M_F(t).\label{eq12}
\end{eqnarray}
The masses are in units of 175 GeV. We repeat for ready reference that 
the 12 fermions are suffixed as $F$ = 1,2,... 12. $F$ = 1,2,3 are the
$U$-quarks: top($t$), charm($c$) and  up($u$). Similarly, $F$=4,5,6 
denote the $D$-quarks: bottom($b$), strange($s$) and down($d$); 
$F$=7,8,9 are the $E$-leptons: electron($e$), muon($\mu$) and tau($\tau$); and
$F$=10,11,12 are the $N$-neutrinoes $\nu_e,  \nu_{\mu}, \nu_{\tau}$.
Further, $M_1=M_{top},~M_2=M_{charm},~\cdots, M_{12}=M_{tau}$.

$A_F$ is a group theoretic factor whose value is `6' for quarks, i.e.,
$F$=1,2,$\cdots$,6 and `4' for the leptons i.e., $F$=7,8,$\cdots$ , 12. The positive
values indicate the field theory containing Yukawa couplings only and may
not be asymptotically free.

$Y_F$ is the mixing term which can be put in matrix form
\begin{equation}
Y_F=\sum_H A_{FH}M_H^\dagger(t) M_H(t),~~~~~~~H=1,2, \cdots , 12.\label{eq13}
\end{equation}
In MSSM, the matrix $A_{FH}$ is specified by the 144 elements given below,
\begin{eqnarray}
A_{FH}=
\left (
\begin{array}{cccccccccccc}
 0 & 3 & 3 & 1 & 0 & 0 & 0 & 0 & 0 & 1 & 1 & 1\\
3 & 0 & 3 & 0 & 1 & 0 & 0 & 0 & 0 & 1 & 1 & 1\\
3 & 3 & 0 & 0 & 0 & 1 & 0 & 0 & 0 & 1 & 1 & 1\\
1 & 0 & 0 & 0 & 3 & 3 & 1 & 1 & 1 & 0 & 0 & 0\\
0 & 1 & 0 & 3 & 0 & 3 & 1 & 1 & 1 & 0 & 0 & 0\\
0 & 0 & 1 & 3 & 3 & 0 & 1 & 1 & 1 & 0 & 0 & 0\\
0 & 0 & 0 & 3 & 3 & 3 & 0 & 1 & 1 & 1 & 0 & 0\\
0 & 0 & 0 & 3 & 3 & 3 & 1 & 0 & 1 & 0 & 1 & 0\\
0 & 0 & 0 & 3 & 3 & 3 & 1 & 1 & 0 & 0 & 0 & 1\\
3 & 3 & 3 & 0 & 0 & 0 & 1 & 0 & 0 & 0 & 1 & 1\\
3 & 3 & 3 & 0 & 0 & 0 & 0 & 1 & 0 & 1 & 0 & 1\\
3 & 3 & 3 & 0 & 0 & 0 & 0 & 0 & 1 & 1 & 1 & 0
\end{array}
\right ).\label{eq14}
\end{eqnarray}
The diagonal elements of $A_{FH}$ have been taken out as the cubic term in
equation(\ref{eq11}); so they are zero.

As the model is minimal supersymmetric, the gauge factors $G_F(t)$, which are the sum
of gauge couplings, are fixed, we take the values from reference\cite{a14} and~\cite{a7}.
\begin{equation}
G_U(t)=\frac{13}{15}g_1^2(t)+3g_2^2(t)+\frac{16}{3}g_3^2(t)=\sum_{i=1}
^{3}K_U^ig_i^2(t).
\end{equation}
$F=$1,2,3 stand for $U$ and they are degenerate electromagnetic gaugewise. Similarly,
\begin{eqnarray}
G_D(t)&=&\frac{7}{15}g_1^2(t)+3g_2^2(t)+\frac{16}{3}g_3^2(t),~~~~F=4,5,6\\
G_E(t)&=&\frac{9}{5}g_1^2(t)+3g_2^2(t),~~~~~~~~~~~~~~~~~~F=7,8,9\\
and~~~~~~~~~~~G_N(t)&=&\frac{3}{5}g_1^2(t)+3g_2^2(t),~~~~~~~~~~~~~~~~~~F=10,11,12.
\end{eqnarray}
Here $K_N^3=K_E^3=0$, as the leptons do not have the strong colour
interaction. We shall need the integrals,
\begin{equation}
-\frac{1}{8\pi^2} \int_{0}^{t}d\tau~G_F(\tau)=\sum_{i=1}^{3}
\frac{K_i^F}{c_i} \log(1-\frac{c_ig_i^2t}{8\pi^2})\label{eq19}
\end{equation}
and
\begin{eqnarray}
-\frac{1}{8\pi^2} \int_{0}^{t_X}d\tau~G_F(\tau)&=&\sum_{i=1}^{3}
\frac{K_i^F}{c_i} \log(1-\frac{c_ig_i^2t_X}{8\pi^2})\label{eq20}\\
 &=&\sum_{i=1}^{3}\frac{K_i^F}{c_i} \log \frac{g_i^2}{g_U^2}.\label{eq21}
\end{eqnarray}
Deo and Maharana\cite{a10} made a very important observation that 
equation(\ref{eq12}), which has to be solved for a given fermion, does not contain
the coeficients of the same mass in the matrix $A$ of equation (\ref{eq14}). 
This fact has been overlooked by all previous authors. As a result, the 
calculational dedails become erroneous and the values obtained are unreliable. The 
present approach gives a hope of a simple method of entangling the mass due to
finding the solution of 12 differentential equations.
As such, the terms $Z_F(t)$ can be exponiented away. We introduce a 
subsidiary mass $m_F(t)$, through
\begin{equation}
M_F(t)=m_F(t) \exp \left(\frac{1}{16\pi^2}\int_0^t Z_F(\tau)~d\tau\right),
\end{equation}
such that $M_F(M_Z)=m_F(M_Z)\equiv m_F(0)$.
They satisfy the equation
\begin{equation}
16\pi^2\frac{dm_F(t)}{m_F^3}=A_F \exp \left( \frac{1}{8\pi^2}\int_0^t
Z_F(\tau)~d\tau \right)~dt.\label{eq23}
\end{equation}
They look, astonishingly, similar to the gauge sector one loop RG equation (\ref{eq9}) and
can be solved exactly. Integrating equation(\ref{eq23}) from $M_Z$ to $M_X$ i.e.,
from $t$=0 to $t_X$, we get 
\begin{equation}
\frac{8\pi^2}{m_F^2(M_Z)}=\frac{8\pi^2}{m_F^2(M_X)}+A_F \int_0^{t_X}dt~
\exp \left( \frac{1}{8\pi^2}\int_0^tZ_F(\tau) d\tau\right).\label{eq24}
\end{equation}
Putting back the exponential,
\begin{equation}
\frac{M_{top}^2(M_Z)}{M_F^2(M_Z)}=\frac{M_{top}^2(M_Z)}{M_F^2(M_X)}
\exp \left( \frac{1}{8\pi^2}\int_0^{t_X}Z_F(\tau) d\tau \right)
+\frac{A_F}{8\pi^2}\int_0^{t_X}dt~
\exp \left( \frac{1}{8\pi^2}\int_0^{t} Z_F(\tau) d\tau\right).\label{eq25}
\end{equation}
This is the exact one loop solution. $M_F^2(M_Z)$ is  the  mass of the fermions at $M_Z$. 

The descent or ascent  running of the masses from GUT $M_X$ to electroweak $M_Z$, can
be obtained by integrating equation(\ref{eq23}) from $t=t_X$ to $t$. The result which
is not given in reference~\cite{a10} is
\begin{equation}
\frac{8\pi^2M_{top}^2}{M_F^2(t)}=8\pi^2\frac{M_{top}^2}{M_F^2(M_X)} \exp
\left( \frac{1}{8\pi^2}\int_t^{t_X}Z_F(\tau)d\tau\right)+A_F\int_t^{t_X}
dt_1 \exp\left( \frac{1}{8\pi^2}\int_t^{t_1}Z_F(\tau)d\tau\right).\label{eq26}
\end{equation}
This is also one loop exact. By solving equations (\ref{eq25}) and (\ref{eq26}), we can
find $M_F(M_X)$ and $M_F(t)$ repsectively.

\section{Original mass of all fermions at $M_X$ ?}
The gauge integrals over $G_F(t)$ are easily and acurately calculable.
The most difficult task is to evaluate $Y_F(t)$. Even though, it does not
contain $M_F(t)$, it is a sum of squares of moduli of the masses of all
fermions. For example, from the matrix as given by equation(\ref{eq14}),
\begin{equation}
Y_{top}(t)=3M_c^2(t)+3M_u^2(t)+M_b^2(t)+M_{\nu_e}^2(t)+M_{\nu_\mu}^2(t)
+M_{\nu_\tau}^2(t).
\end{equation}
There is mixing of six other fermions for the top.

In the `top heavy integral' approximation, one retains only those terms containing 
$M_{top}^2(t)$ occuring in any integral with $Y_F(t)$ . Consider the top case.  
$Y_{top}(t)$ can be set equal to zero in the RG Equation for the top. The top mass 
is then  given by a simple expression
\begin{equation}
1= \frac{M_{top}^2(M_Z)}{M_{top}^2(M_X)}C_{top}+D_{top},
\end{equation}
where, using integrals (\ref{eq19}) to (\ref{eq21}),
\begin{eqnarray}
C_{top}&=&\prod_{i=1}^{3}{ \left( \frac{g_i^2}{g_U^2}\right)}^{\frac{K_i^U}
{c_i}}=0.086\\
\text{and}\\
D_{top}&=&\frac{6}{8\pi^2}\int_0^{t_X=33}dt \prod_{i=1}^3{\left( 1-
\frac{g_i^2c_it}{8\pi^2}\right)}^{\frac{K_i^U}{c_i}}=0.802
\end{eqnarray}
Putting these values, the original mass of the top, at an energy of
$2.2\times10^{16}$ GeV was, approximately,
\begin{equation}
M_{top}(M_X)=M_{top}(M_Z){\left(1-D_{top}\right)}^{-1/2}C_{top}^{1/2}
=114~~ \text{GeV}.
\end{equation}
This is Deo-Maharana result.

We can write a general formula for all fermions
\begin{equation}
\frac{M_{top}^2(M_Z)}{M_F^2(M_Z)}=\frac{M_{top}^2(M_Z)}{M_F^2(M_X)}C_F
+D_F,\label{eq32}
\end{equation}
where
\begin{eqnarray}
C_F&=&\prod_{i=1}^{3}{\left(\frac{g_i^2}{g_U^2}\right)}^{\frac{K_i^F}{C_i}}
\exp\left(\frac{1}{8\pi^2}\int_0^{t_X}Y_F(\tau)d\tau\right)\label{eq33}\\
and ~~~~~~D_F&=&\frac{A_F}{8\pi^2}\int_0^{t_X}dt\prod_{i=1}^{3}{\left(1-\frac{c_{i}g_{i}^{2}t}
{8\pi^2}\right)}^{\frac{K_i^F}{c_i}}\exp\left(\frac{1}{8\pi^2}
\int_0^{t}Y_F(\tau)d\tau\right).\label{eq34}
\end{eqnarray}
It is not so easy to calculate $Y_F$ for other fermions even with 'heavy top integral' 
approximation given by equation (\ref{e6}). So we consider the 
Yukawa-like coupling $h_F(t)$ for $F$=2,3, ..., 12, which is related to $M_F(t)$ as
\begin{equation}
M_F(t)=h_F(t)\exp\left(-\frac{1}{16\pi^2}\int_0^tG_F(\tau)d\tau\right).
\end{equation}
The RG Equation for $h_F(t)$ is,
\begin{equation}
8\pi^2d\left(\log h_F^2(t)\right)=A_Fh_F^2(t)\exp\left(-\frac{1}{8\pi^2}\int_0^tG_F
(\tau)d\tau\right)dt+Y_F(t)dt.\label{eq36}
\end{equation}
Here $h_F(t)$ contains the usual $\beta$-angle factors of the two Higgs system.
As has been discussed in Section-1, we may not need this angle in our approach 
to the problem at hand. If $F>1$, the heaviest fermion next to the top is 
the bottom. The first term of equation(\ref{eq36}), at the $M_Z$ scale, is 
$1/1235$ times smaller, whereas the last term contains one
$h_{top}$ in $h_{botm}$. To a good approximation, and to begin with, we shall neglect
this term and express $h_F(t)$ in terms of $Y_F(t)$'s. Then
\begin{equation}
\int_0^{t_X}Y_F(\tau)d\tau=8\pi^2\log \frac{h_F^2(M_X)}{h_F^2(M_Z)}.
\end{equation}
Here,~~$h_F^2(M_Z)=M_F^2(M_Z)$ by definition. In the units of top mass 175 GeV,
\begin{eqnarray}
h_F^2(M_X)=M_F^2(M_X)\exp \left(\frac{1}{8\pi^2}\int_0^{t_{X}}G_F(\tau)d\tau\right)
&=&M_{top}^2(M_X)\exp \left(\frac{1}{8\pi^2}\int_0^{t_{X}}G_F(\tau)d\tau\right)\\
&=&M_{top}^2(M_Z)=1.
\end{eqnarray}
So,
\begin{eqnarray}
\int_0^{t_X}Y_F(\tau)d\tau&=&-8\pi^2\log M_F^2(M_Z)\\
or~~~~~~ \exp \left(\frac{1}{8\pi^2}\int_0^{t_{X}}Y_F(\tau)d\tau\right)
&=&\frac{M_{top}^2(M_Z)}{M_{F}^2(M_Z)}.\label{e64}
\end{eqnarray}
This is true as long as the first term of equation(\ref{eq36}) is negligible. Since
there is $M_F^3$ in this term, it is much more justifiable to set 
$D_{lepton}=0$. The general equation (\ref{eq32}) for the leptons gives
\begin{equation}
\frac{M_{top}^2(M_Z)}{M_{lepton}^2(M_Z)}=\frac{M_{top}^2(M_Z)}
{M_{lepton}^2(M_X)}C_{lepton},
\end{equation}
where
$$
C_{lepton}=\prod_{i=1}^{3}{\left(\frac{g_i^2}{g_U^2}\right)}^{\frac{K_i^{lepton}}{C_i}},
$$
\begin{eqnarray}
M_{lepton}^2(M_X)&=& M_{lepton}^2C_{lepton}\\
  \text{or}~~~~~~~~M_{lepton}(M_X)&=& M_{lepton}^{expt.}C_{lepton}^{1/2}.
\end{eqnarray}
Putting the gauge constants  and masses in the above, as experimentally reported, 
we get the masses at GUT scale for different leptons as given in Table-\ref{tab:table1},

\begin{center}
\begin{table}[h]
\caption{\label{tab:table1}Unification scale mass for leptons}
\begin{tabular}{|c|c|}\hline
Lepton & Unification scale mass (GeV)\\
\hline
e& 116\\
$\mu$ & 116\\
$\tau$ & 116\\
$\nu_{e}$ & 126\\
$\nu_{\mu}$ & 126\\
$\nu_{\tau}$ & 126\\ \hline
\end{tabular} 
\end{table}
\end{center}
At the grand unification mass, the electron, the muon and the tau climb to 116 GeV 
in this approximation.
The calculation for the neutrinoes are not reliable as the isospin factors are
uncertain and may be inaccurate.

We are now left with the five quarks. For them, we attempt to find the next leading order
approximation,  i.e., we first set the first term equal to zero and obtain
\begin{equation}
\exp \left(\frac{1}{8\pi^2}\int_0^{t_{X}}Y_Q(\tau)d\tau\right)
=\frac{1}{h_Q^2(M_Z)}=\frac{M_{top}^2(M_Z)}{M_{Q}^2(M_Z)}.
\end{equation}
This is used in the calculation for $C_Q$ of equation(\ref{eq33}), which gives
\begin{equation}
C_Q=\prod_{i=1}^{2}{\left(\frac{g_i^2}{g_U^2}\right)}^{\frac{K_i^Q}{C_i}}
\frac{1}{M_Q^2(M_Z)}.
\end{equation}
Using
\begin{equation}
\exp \left(\frac{1}{8\pi^2}\int_0^{t}Y_F(\tau)d\tau\right)
=\frac{h_Q^2(t)}{M_{Q}^2(M_Z)},\label{eq47}
\end{equation}
in equation(\ref{eq34}), we get
\begin{equation}
D_Q=\frac{6}{8\pi^2}\int_0^{t_X}\frac{h_Q^2(t)}{M_Q^2(M_Z)}\prod _{i=1}
^{2}{\left(1-\frac{g_{i}^{2}c_{i}t}{8\pi^2}\right)}^{\frac{K_{i}^{Q}}{c_i}}dt.
\end{equation}
We look for ways to calculate $h_Q(t)$. We can retain only the top quark 
in the RG equation in a slightly different way than what has
been taken by RRR\cite{a6} and get,
\begin{equation}
8\pi^2 d \log h_Q^2(t)\simeq N_Q dt.
\end{equation}
We have
\begin{equation}
N_{charm}=3,~~~  N_{up}=3, ~~~N_{bottom}=1, ~~~N_{strange}=N_{down}=0.
\end{equation}
 To use this value in $Y_F$, it is necessary to take an average as the couplings
change very rapidly,
\begin{equation}
\overline{\log h_Q^2(t)}=\frac{1}{t_X}\int_0^t \frac{d}{d\tau} \log
 h_Q(\tau)d\tau =\frac{1}{t_X}\int_0^t \frac{N_Q}{8\pi^2}d\tau
=\frac{t}{t_X}\frac{N_Q}{8\pi^2},
\end{equation}
or
\begin{equation}
\overline{h_Q^2(t)}=\exp\left(\frac{t}{t_X}\frac{N_Q}{8\pi^2}\right).\label{e75}
\end{equation}
This expression for Yukawa coupling, is not much different from unity  for all 
allowed $t$'s. We arrive at the following result,
\begin{equation}
D_Q=\frac{6}{8\pi^2}\int_0^{t_X}dt\exp\left(\frac{t}{t_X}\frac{N_Q}{8\pi^2}\right)
\prod _{i=1}
^{3}{\left(1-\frac{g_{i}^{2}c_{i}t}{8\pi^2}\right)}^{\frac{K_{i}^{Q}}{c_i}}.
\end{equation}
The unification mass is calculated numerically from
\begin{equation}
M_Q(M_X)=M_{top}{\left(\frac{C_Q}{1-D_Q}\right)}^{1/2}.
\end{equation}
The results, for the quarks, are given in Table-\ref{tab:table2}. We note that
$M_{charm}(M_X)=M_{up}(M_X)\neq M_{top}(M_X)$ and
$M_{strange}(M_X)=M_{down}(M_X)\neq M_{bottom}(M_X)$.
\begin{center}
\begin{table}[h]
\caption{\label{tab:table2}Unification scale mass for quarks}
\begin{tabular}{|c|c|}\hline
Quark & Unification scale mass (GeV)\\
\hline
Top & 114\\
Charm & 115\\
Up & 115\\
Bottom & 119\\
Strange & 118\\
Down & 118\\ \hline
\end{tabular}
\end{table}
\end{center}
Thus, we have shown that all fermions seem to originate at an energy
$2.2 \times 10^{16}$ GeV with equal mass of about 115 GeV. Perhaps, this is
due to the equality of A$\simeq$ 1 in Wolfenstein's parametrisation of CKM matrix.
In this perturbative method of solution for finding the unification mass, 
information about the masses of the 11 fermions has been lost due to cancellation
of $M^2(M_Z)$ in both r.h.s. and l.h.s. of the equation (\ref{eq25}) due to use of 
equations (\ref{e64}) and (\ref{eq47}). The result of a common mass at the origin
is atleast a hypothesis and has an approximation~\cite{a10}.

The top mass decreases with energy but 
all the other quarks and leptons, starting from the value at $M_X$ acquire smaller and
smaller values and become quite light in the electroweak scale. The descent 
or ascent equation(\ref{eq26}) describing the `run' can be put in a form like 
equation(\ref{eq32}).
\begin{equation}
\frac{M_{top}^2(M_Z)}{M_{F}^2(t)}=\frac{M_{top}^2(M_Z)}{M_{F}^2(M_X)}C_F(t)
+D_F(t),\label{eq55}
\end{equation}
where
\begin{equation}
C_F(t)=a_F(t) \exp\left(\frac{1}{8\pi^2}\int_0^{t_X}Y_F(\tau)d\tau\right),\label{eq56}
\end{equation}
\begin{equation}
D_F(t)=\frac{A_F}{8\pi^2}\int_t^{t_X}dt_1 b_{F}(t_1)
\exp\left(\frac{1}{8\pi^2}\int_t^{t_1}Y_F(\tau)d\tau\right),\label{eq57}
\end{equation}
\begin{equation}
a_F(t)=\prod_{i=1}^{3} {\left[\left(1-\frac{g_i^{2} c_{i} t_{X}}
{8\pi^2}\right)\over \left(1-\frac{g_i^{2} c_{i} t}{8\pi^2}\right)\right]}
^{\frac{K_i^F}{c_i}},\label{eq58}
\end{equation}
\begin{equation}
and ~~~~b_F(t)=\prod_{i=1}^{3}{\left[\left(1-\frac{g_i^{2} c_{i} t_{1}}
{8\pi^2}\right)\over \left(1-\frac{g_i^{2} c_{i} t}{8\pi^2}\right)\right]}
^{\frac{K_i^F}{c_i}}.\label{eq59}
\end{equation}
For the top, we can take $Y_F\rightarrow0$ and calculate variation of
its mass from $M_U\simeq 115$ GeV to the top mass 175 GeV.
\begin{equation}
M_{top}(t)=\frac {M_{top}(M_X)} {{\left( C_{top}(t)+\frac{M_U^2}{M_{top}^2}D_{top}(t)
\right)}^{1/2}}\label{eq60}
\end{equation}
The values are given in Table-\ref{tab:table5} of Section-8.

For other cases, we shall try to fit them into a scheme which is
not only much simpler, but has better physical content. 
However, in the following section, we discuss a different route 
for solutions without specifying gauge factors completely.

\section{Deduction of Wolfenstein parameters and the rotational integers}
Let us construct a function $B_F(t)$ such that 
\begin{equation}
8\pi^2\frac{d}{dt}\log B_F^2(t)=Z_F(t).
\end{equation}
Then
\begin{eqnarray}
\exp\left(\frac{1}{8\pi^2}\int_0^{t_X} Z_F(\tau)d\tau\right)
&=&\frac{B_F^2(M_X)}{B_F^2(M_Z)},\\
\exp\left(\frac{1}{8\pi^2}\int_{t_X}^0 Z_F(\tau)d\tau\right)
&=&\frac{B_F^2(M_Z)}{B_F^2(M_X)},\\
\exp\left(\frac{1}{8\pi^2}\int_0^{t} Z_F(\tau)d\tau\right)
&=&\frac{B_F^2(t)}{B_F^2(M_Z)},\\
\text{and} \exp\left(\frac{1}{8\pi^2}\int_{t_X}^t Z_F(\tau)d\tau\right)
&=&\frac{B_F^2(t)}{B_F^2(M_X)}.
\end{eqnarray}
Equation(\ref{eq25}) reduces to
\begin{equation}
M_F(M_Z)=\frac{\frac{B_F(M_Z)}{B_F(M_X)}M_F(M_X)}{{\left(1+\frac{M_F^2(M_X)}{M_{top}^2}
 \frac{A_F}{8\pi^2}\int_0^{t_X}dt \frac{B_F^2(t)}{B_F^2(M_Z)}  \right)}^{1/2}}.
\end{equation}

 As indicated in the last section, except the top, we neglect the terms with $A_F$. Then
\begin{equation}
M_F(M_Z) \simeq M_F(M_X)\frac{B_F(M_Z)}{B_F(M_X)}=M_{F}(M_X) \exp\left(-\frac{I_F}{16\pi^2}
\right)= M_{F}(M_X)\lambda^{n_F} ,\label{eq67}
\end{equation}
where the integral $I_F$ is
\begin{equation}
I_F=\int_0^{t_X}Z_F(t)dt=\int_0^{t_X}\left(\sum_G A_{FG}M_G^{\dagger}(t)M_G(t)
-G_F(t)\right)dt,\label{eq68}
\end{equation}
with $A_{1G}$=0 to indicate that this is not for the top. Since the ratio of the 
two fermion masses can be put as $\frac{M_{F_1}}{M_{F_2}}=\lambda$,
the parameter $M_{\nu}$ cancels out and $\lambda$ is the Wolfenstein parameter.
The purpose of introducing the logarithms is to incorporate the possible
multivaluedness of $I_F$ and obtain the integral powers of $\lambda$.
From equation (\ref{eq68}),
\begin{eqnarray}
I_F&=& \int_0^{t_X}Z_F(t)dt=\int_0^{t_X}\left(\sum_G A_{FG}M_G^{\dagger}(t)M_G(t)
-G_F(t)\right)dt\label{eq69}\\
&=&\frac{1}{2}\int_0^{t_X}\left(\sum_G A_{FG}M_G^{\dagger}(t)M_G(t)
-G_F(t)   +\sum_G A_{FG}M_G^{\dagger}(-t)M_G(-t)-G_F(-t)\right)dt\\
&=&\frac{1}{4}\int_{-t_X}^{t_X} \frac{dt}{dM_F}dM_F(t)
\left(\sum_G A_{FG}M_G^{\dagger}(t)M_G(t)
-G_F(t)   +\sum_G A_{FG}M_G^{\dagger}(-t)M_G(-t)-G_F(-t)\right)\\
&=& \frac{16\pi^2}{4}\int_{-M_U}^{M_U} \frac{dM_F(t)}{M_F(t)}
\frac{\left (\sum_G A_{FG}M_G^{\dagger}(t)M_G(t)-\frac{1}{2}(G_F(t)+G_F(-t))\right )} 
{\left( A_FM_F^{\dagger}(t)M_F(t)-G_F(t)\right )},\label{eq68a}
\end{eqnarray}
where we have used equation (\ref{eq12}). 
Let us set
\[
M_F(t)=M_U e^{i\theta_F(M_F(t))n_F}= M_U e^{i\theta_F(t)n_F},
\]
where $n_F$ is an integer and we will call it rotational integer. 
Using the above in (\ref{eq68a}), we get
\begin{equation}
I_F=in_F\frac{16\pi^2}{2}\int_{-t_X}^{t_X}d\theta_F(t)
\frac{\sum_G A_{FG}}{A_F+{\sum_G A_{FG}-\frac{G_F(t)}{M_U^2}}}.\label{eq68b} 
\end{equation}

We have omitted inconsequential factor $\frac{G_F}{M_U^2}$ in the numerator. We note that
for quarks $A_F+\sum_{G=1}^6$=6+7=13 and for leptons 4+9=13. The integral is almost
a constant and isolate the integer $n_F$, characterising F from the inegral. Using
$1=\frac{1}{12}\sum_H$ in equation (\ref{eq68b}), we have
\begin{equation}
I_F=i n_F\frac{16\pi^2}{2}\frac{1}{12}\sum_H\sum_{G}A_{HG}\int_{-t_X}^{t_X} 
d\theta_H(t)\frac{1}{A_H+\sum_{G}A_{HG}-\frac{G_H}{M_U^2}}.\label{eq68c}
\end{equation}
In the above we have averaged over the twelve fermions.
Retracing the steps and using $M_H(t)=M_Ue^{i\theta_H(t)M_U^2}$ in equation(\ref{eq68c}), 
 we finally get
\begin{equation}
I_F\simeq n_F\frac{1}{12}\sum_H\sum_G A_{HG}
\int _0^{t_X}dt=\frac{1}{12}n_Ft_X\sum_H\sum_G A_{HG}.\label{eq98}
\end{equation}
First, we shall be interested in the CKM matrix for quarks with lepton masses 
taken as zero. Then let G vary from 1 to 6, whereas F will be taking values from 2 to 12, 
since all of them contains 6 quarks. Only the coefficients
$A_{HG}$  are needed to calculate $I_F$ of equation (\ref{eq98}). 
The coefficients $A_{FG}$ have non-vanishing values for $F$=2,3,$\cdots$,12 and 
$G$=1,2,$\cdots$ ,6.
\begin{equation}
 \text{For}~~~ F= 2, \cdots, 6:~~~~~~~~~~  A_{F1}+A_{F2}+ \cdots +A_{F6}=7,
\end{equation}
and
\begin{equation}
\text{For} ~F= 7, \cdots ,12:~~~~~~~~~~ A_{F1}+A_{F2}+ \cdots +A_{F6}=9.
\end{equation}
For the twelve fermions, the sum of the values of the coefficients $A_{FG}$
is $(7\times5)+(9\times6)=89$.
The average of Z, i.e., $\bar{Z}$ is geven
\begin{equation}
\bar{Z}=89/12,~~~~~~~~~~~~~\text{and}~~~~~~~~~~~~~~~ I_F=n_F~ t_X ~\frac{89}{12}.
\end{equation}
From equation(\ref{eq67}), the Wolfenstein parameter $\lambda$ turns out to be 
\begin{equation}
\lambda=\exp\left(-\frac{t_X}{16\pi^2} \frac{89}{12}\right)=0.219
\end{equation}
This is an excellent result in spite of the approximate estimates. The masses
of all the fermions due to quark-lepton equivalence other than the top is 
\begin{equation}
M_F(M_Z)\simeq M_{F}(M_X)\lambda^{n_F}\simeq \lambda^{n_F}M_U\label{eq80a}
\end{equation}
The table-\ref{tab:table3} identifies the particles. We have increased $n_F$ by
neighbourhood integers which we have called the rotational integers.
\begin{center}
\begin{table}[h]
\caption{\label{tab:table3}Identification of fermions}
\begin{tabular}{|c|c|c|}\hline
$n_F$&Mass in GeV & Fermion\\
\hline
2 &5.5&bottom(b)\\
3 &1.2& charm(c),tau lepton($\tau$)\\
4 &0.264&strange(s), muon($\mu$)\\
6 &0.012&down(d),tau neutrino($\nu_{\tau}$)\\
7 &0.0028&up(u)\\
8 & $6\times 10^{-4}$ & electron(e)\\ 
9 & $1.33 \times 10^{-4}$ & muon neutrino($\nu_{\mu}$)\\
16& $3\times 10^{-9}$& electron neutrino($\nu_e$)\\\hline
\end{tabular}
\end{table}
\end{center}

\section{Gauge contributions}
\subsection{The top mass}
It is easy to write down the equation for the top mass in terms of the mass
$M_F(M_X)$ and gauge couplings.
\begin{equation}
M_{top}=M_{top}(M_X){\left(\frac{1-d_{top}}{a_{top}} \right)}^{1/2}.
\end{equation}
We shall also need the equation for the descent of top mass
\begin{equation}
\frac{M_{top}^2}{M_{top}^2(t)}=\frac{M_{top}^2}{M_{top}^2(M_X)}a_{top}(t)+ D_{top}(t).
\end{equation}
The equation contains only the gauge factors.

\subsection{ The masses of charm and up quarks.}
The RG Equation for the top is non-linear and intermixed. Specifically it is
\begin{equation}
8\pi^2\frac{d}{dt}\log M_{top}^2(t)=-G_U+6M_{top}^2(t)+Y_{top}(t).
\end{equation}
Following a texture analysis of the RG solutions, we introduce a function
$B(t)$,( not to be confused with $B_F(t)$), which satisfies
\begin{equation}
8\pi^2\frac{d}{dt}\log B^2(t)=-G_U+6M_{top}^2(t).\label{eq83}
\end{equation}
Equation(\ref{eq83}) does not specify the function $B(t)$ completely except that
$d \log B(t)= d \log(M_{top}(t))$. This lone restriction gives us an infinite
number of free choices. Because letting B$\rightarrow \xi$B, where $\xi$ is an 
arbitrary constant, does not change the top mass function $M_{top}(t)$. 
For simplicity, we take the second derivative of this function $B(t)$ to be zero so that
\begin{equation}
d \log B(t) =C_B dt.
\end{equation}
$C_B$ is essentially $d \log M_{top}(t)$, $M_{top}(t)$ changes by only 30 to 50 GeV
as the mass $ \mu$ goes from 91 GeV to $ 2.2 \times 10^{16}$ GeV.
$C_B \simeq (175-(123~~\text{ to}~~ 115))/175 \simeq 0.297~~\text{ to}~~ 0.342$. We take
$C_B = 0.3$.
The nonequiness due to nonlinearity can also be seen as follows:
\[
dlogB(t)= dlogM(t)=dlog\frac{M_{top}}{M_o}=\frac{dt}{16\pi^2}
\left [ -G_U+6\left (\frac{M_{top}}{M_o}\right )^2  \right ].
\]
$M_o$ is an arbitrary constant. We can choose $M_o$ suitably so that
\[
\left [ -G_U+6 \left (\frac{M_{top}}{M_o}\right )^2\right]/16\pi^2\simeq 0.3=C_B.
\]
Integrating from $t=t_X(M_X)$ to $t=0 (M_Z)$, we have
\begin{equation}
\log \frac{B_Z}{B_U}=-C_B t_X,\label{eq85}
\end{equation}
and
\begin{equation}
\frac{B_Z}{B_U}=e^{-C_B t_X}=(0.192)^6.
\end{equation}
Furthermore, integrating from $t=t_X$ to arbitrary $t$,
\begin{equation}
\log \frac{B(t)}{B_U}=C_B(t-t_X),
\end{equation}
or
\begin{equation}
\frac{B(t)}{B_U}=\exp \left( C_B(t-t_X) \right)=\left[\left( \frac{B_Z}{B_u}\right)^{1/6}
\right]^{(1-t/t_X)}.
\end{equation}
Let us examine the case for the charm. We have 
\begin{eqnarray}
8\pi^2\frac{d}{dt} \log M_c^2(t)&=&-G_U+3M_t^2,\nonumber \\
&=&-G_U+\frac{1}{2}[6M_t^2-G_U]+\frac{1}{2}G_U,\nonumber \\
&=&-\frac{1}{2}G_U+4\pi^2\frac{d}{dt} \log B^{2}(t).
\end{eqnarray}
Integrating from $t_X$ to $t$,
\begin{eqnarray}
\log \frac{M_{charm}^2(t)}{M_{charm}^2(M_X)} &=&\frac{1}{2} \log \frac{B^2(t)}{B_U^2}
-\frac{1}{16\pi^2} \int_{t_X}^t G_U(t)dt,\\
M_{charm}(t)&=&M_U{\left(\frac {B(t)}{B_U}\right)}^{1/2} \exp \left(-\frac{1}{32\pi^2}
\int_{t_X}^t G_U(\tau)d\tau\right),\\
M_{charm}(M_Z)&=&M_U{\left(\frac {B_Z}{B_U}\right)}^{1/2} C_{top}^{-1/4},~~~~~
C_{top}^{-1/4}=1.8466,\\
and ~~~~M_c&=& {\left[{ \left(\frac {B_Z}{B_U}\right)}^{1/6} C_{top}^{-1/12}\right]}^3 
M_{charm}(M_X).\label{eq100a}
\end{eqnarray}
From section-4, we now calculate
\begin{equation}
\lambda_{charm}=e^{-c_B t_{X}/6}C_{top}^{-1/12}=0.221
\end{equation}
Again this is a good result. The mass of the charm is 1.24 GeV from the equation (\ref{eq100a})
in very good agreement 
with the experimental value. The rotational integer $n_F$ is three as in Table-\ref{tab:table3}.

We continue further and consider the up quark. The `heavy top integral' approximation 
of the RGE for the up quark is
\begin{equation}
8\pi^2 \frac{d}{dt} \log M_{up}^2=-G_U+3M_{top}^2(t).
\end{equation}
Guessing from the general rotational integral parametrization, as shown in 
Table-\ref{tab:table3} as $\lambda^7$, we write this equation as 
\begin{eqnarray}
8\pi^2 \frac{d}{dt} \log M_{up}^2&=&-G_U +\frac{1}{2}(6M_{top}^2(t)),\nonumber\\
&=&-\frac{G_U}{2}+8\pi^2 \frac{d}{dt} \log B^2(t)-\frac{1}{2}
8\pi^2 \frac{d}{dt} \log M_{top}^2(t).
\end{eqnarray}
Integrating, we get
\begin{eqnarray}
\frac{M_{up}}{M_U}&=&\left( \frac{B_Z}{B_u}\right){\left(\frac{M_U}{M_{top}}
 \right)}^{1/2}a_U^{-1/4},\\
{\lambda}_{up}&=&\left[\left( \frac{B_Z}{B_u}\right){\left(\frac
{M_U}{M_{top}}\right)}^{1/2}a_U^{-1/4} \right]^{1/7}=0.220.
\end{eqnarray}

This gives $M_{up} \simeq$ 0.0029 GeV, with $M_U=M_{top}(M_X)$, a little lower 
value but quite close to the experimental result. The values
depend crucially on $C_B$, which is itself not exact. It is also true that $M_U$
may be a little different from 115 GeV.

\section{ Masses of the down quarks}
The calculation of the mass of up-quark has set the method of finding solutions
close to the experimental values by using the texture analysis function $B(t)$
ignoring the terms with coefficients $A_F$. We present a general recipe from the 
`heavy top integral' approximation. Let the RG Equation for any $F \neq 1 $ be
\begin{equation}
8\pi^2 \frac{d}{dt} \log M_F^2=-G_F+N_F M_{top}^2.
\end{equation}
Here $N_F$ can be zero as well.
\begin{eqnarray}
N_F M_{top}^2(t)&=&\frac{N_F}{6} \left( 6M_{top}^2(t)-G_U \right) 
+\frac{N_F}{6} G_U,
\nonumber \\
&=&\frac{N_F+\eta _F}{6}\left(8\pi^2 \frac{d}{dt} \log B^2(t) \right)
-\frac{\eta _F}{6}\left(8\pi^2 \frac{d}{dt} \log M_{top}^2(t) \right)+\frac{N_F}{6} G_U .
\end{eqnarray}
$\eta_F$ can be any arbitrary coefficient. All of them will satisfy the 
RG Equation in the `heavy top integral' approximation. But they should let $ n_F$ and $\lambda$ 
be such that they are within rotational integers and gauge interaction contributions.

Integrating from the known values, $t_X$ to 0, we get
\begin{equation}
M_F(M_Z)=M_U \exp \left(\frac{1}{16\pi^2} \int_0^{t_X}G_F(t)~dt 
-\frac{N_F}{96\pi^2}\int_0^{t_X}G_U(t)~dt\right )\left(\frac{B_Z}{B_U} \right)
^{\frac{N_F+\eta _F}{6}}\left(\frac{M_U}{M_{top}} \right)^{\frac{\eta _F}{6}}.
\end{equation} 
For the down quarks,
\begin{eqnarray}
8\pi^2 \frac{d}{dt} \log M_b^2&=&-G_D+M_{top}^2,\\
8\pi^2 \frac{d}{dt} \log M_s^2&=&-G_D,\\
8\pi^2 \frac{d}{dt} \log M_d^2&=&-G_D,\\
\text{So}~~~~~~~~~8\pi^2 \frac{d}{dt} \log M_b^2&=&-G_D+2M_{top}^2-M_{top}^2,\\
&=&-G_D+\frac{2.5}{6}(6M_{top}^2-G_U)-\frac{1.5}{6}(6M_{top}^2-G_U)+\frac{G_U}{6},\\
&=&-G_D+\frac{2.5}{6}(8\pi^2 \frac{d}{dt} \log B^2)-\frac{1.5}{6}
(8\pi^2 \frac{d}{dt} \log M^2_{top})+\frac{G_U}{6}.
\end{eqnarray}
Integrating from $t_X$ to 0, we get 
\begin{eqnarray}
\frac{M_b}{M_U}&=&{\left(\frac{B_Z}{B_U} \right)}^{2.5/6}{\left(\frac{M_U}{M_{top}} 
\right)}^{1.5/6}a_D^{-1/2}a_U^{1/12},\\
&=&{\left[{\left(\frac{B_Z}{B_U} \right)}^{2.5/12}
{\left(\frac{M_U}{M_{top}}\right)}^{1.5/12}a_D^{-1/4}a_U^{1/24} \right]}^2.
\end{eqnarray}
The quantity in the square bracket is $\lambda \simeq$ 0.203. This gives the value of the 
bottom mass as to be  4.7 GeV.

For the strange we have,
\begin{equation}
8\pi^2 \frac{d}{dt} \log M_s^2=-G_D+\frac{4.5}{6}\left(6\pi^2 \frac{d}{dt}
 \log B^2-G_U\right)-\frac{4.5}{6}\left(6\pi^2 \frac{d}{dt}
 \log M_{top}^2-{G_U}\right).\label{e104}
\end{equation}
This  leads to
\begin{equation}
\frac{M_s}{M_U}={\left(\frac{B_Z}{B_U} \right)}^{4.5/6}{\left(\frac{M_U}{M_{top}}
\right)}^{4.5/6}a_D^{-1/2}=\left[ \left(\frac{B_Z}{B_U} \right)^{4.5/24}{\left
(\frac{M_U}{M_{top}}\right)}^{4.5/24}a_D^{-1/8}\right]^4 = \lambda^4.\label{e105}
\end{equation} 
Here $ \lambda$ comes out to be near 0.196 and the strange mass is found to be 0.168 GeV.

Proceeding further to the down quark, we get
\begin{equation}
8\pi^2 \frac{d}{dt} \log M_d^2=-G_D+\frac{6.5}{6}\left(6\pi^2 \frac{d}{dt}
\log B^2-G_U\right)-\frac{6.5}{6}\left(6\pi^2 \frac{d}{dt}
\log M_{top}^2-{G_U}\right),\label{e106}
\end{equation}
which gives
\begin{equation}
\frac{M_d}{M_U}={\left(\frac{B_Z}{B_U} \right)}^{6.5/6}{\left(\frac{M_U}{M_{top}}
\right)}^{6.5/6}a_D^{-1/2}=\left[ \left(\frac{B_Z}{B_U} \right)^{6.5/36}{\left
(\frac{M_U}{M_{top}}\right)}^{6.5/36}a_D^{-1/12}\right]^6 = \lambda^6.\label{e107}
\end{equation}
This gives  $\lambda\simeq 0.19$ and  $M_s$=0.0053 GeV. The ratio $M_s /M_d \simeq$ 31.
 However, from equation(\ref{eq8}), this ratio is
\begin{equation}
\frac{M_s}{M_d}=\lambda^{-2}=(0.2)^{-2}=25.\label{e108}
\end{equation}
These can be considered as gauge interaction corrections to the Wolfenstein parameter.

\section{The masses of the leptons}
Ignoring the non-linear terms proportional to $A_F$=4, the RG Equation, retaining the 
top quark only, for the six leptons are 
\begin{eqnarray}
8\pi^2 \frac{d}{dt} \log M_e^2&=&-G_E,\\
8\pi^2 \frac{d}{dt} \log M_{\mu}^2&=&-G_E,\\
8\pi^2 \frac{d}{dt} \log M_{\tau}^2&=&-G_E,\\
8\pi^2 \frac{d}{dt} \log M_{\nu_{e}}^2&=&-G_N+3M_{top}^2,\\
8\pi^2 \frac{d}{dt} \log M_{\nu_{\mu}}^2&=&-G_N+3M_{top}^2,\\
8\pi^2 \frac{d}{dt} \log M_{\nu_{\tau}}^2&=&-G_N+3M_{top}^2.
\end{eqnarray}
For the first three electron-leptons, we write, with choices based on $\lambda \simeq .22$
and rotational integer $n_F$,
\begin{eqnarray}
8\pi^2 \frac{d}{dt} \log M_e^2&=&-G_E+\frac{30}{24}\left(8\pi^2 \frac{d}{dt} \log
B^2(t)-G_U\right)-\frac{30}{24}\left(8\pi^2 \frac{d}{dt} \log M_{top}^2(t)-G_U\right),\\
8\pi^2 \frac{d}{dt} \log M_{\mu}^2&=&-G_E+\frac{17}{24}\left(8\pi^2 \frac{d}{dt} \log
B^2(t)-G_U\right)-\frac{17}{24}\left(8\pi^2 \frac{d}{dt} \log M_{top}^2(t)-G_U\right),\\
8\pi^2 \frac{d}{dt} \log M_{\tau}^2&=&-G_E+\frac{11}{24}\left(8\pi^2 \frac{d}{dt} \log
B^2(t)-G_U\right)-\frac{11}{24}\left(8\pi^2 \frac{d}{dt} \log M_{top}^2(t)-G_U\right),
\end{eqnarray}
and obtain
\begin{equation}
\frac{M_e}{M_U}={\left(\frac{B_Z}{B_U} \right)}^{5/4}{\left(\frac{M_U}{M_{top}}
\right)}^{5/4}a_E^{-1/2}=\left[ \left(\frac{B_Z}{B_U} \right)^{5/32}{\left
(\frac{M_U}{M_{top}}\right)}^{5/32}a_E^{-1/16}\right]^8 = \lambda_e^8,\label{e118}
\end{equation}
which gives $\lambda_e=0.209$;  $M_e=4\times10^{-4}$ GeV. Similarly
\begin{equation}
\frac{M_{\mu}}{M_U}={\left(\frac{B_Z}{B_U} \right)}^{17/24}{\left(\frac{M_U}{M_{top}}
\right)}^{17/24}a_E^{-1/2}=\left[ \left(\frac{B_Z}{B_U} \right)^{17/120}{\left
(\frac{M_U}{M_{top}}\right)}^{17/120}a_E^{-1/10}\right]^5 = \lambda_{\mu}^5,\label{e119}
\end{equation}
which gives $\lambda_{\mu}=0.238$;  $M_{\mu}=$ 0.09 GeV,\\
and\\
\begin{equation}
\frac{M_{\tau}}{M_U}={\left(\frac{B_Z}{B_U} \right)}^{11/24}{\left(\frac{M_U}{M_{top}}
\right)}^{11/24}a_E^{-1/2}=\left[ \left(\frac{B_Z}{B_U} \right)^{11/72}{\left
(\frac{M_U}{M_{top}}\right)}^{11/72}a_E^{-1/6}\right]^3 = \lambda_{\tau}^3,\label{e120}
\end{equation}
which gives $\lambda_{\tau}=0.23$;  $M_{\tau}=$ 1.36 GeV.

The masses of the neutrinoes have not yet been measured, only limits have
been set. The reported values, as shown in Table-\ref{tab:table3}, fall into a good pattern,
namely, $M_{\nu _{e}}=M_U {\lambda}^{16}$, $M_{\nu _{\mu}}=M_U {\lambda}^{9}$ 
and $M_{\nu _{\tau}}=M_U {\lambda}^{6}$, $\lambda$=0.22. So the equivalent, 
degeneracy lifting equations, showing the choices of $n_F$ are
\begin{eqnarray}
8\pi^2 \frac{d}{dt} \log M_{\nu_{e}}^2&=&-G_N+\frac{5}{2}\left(6M_{top}^2-G_U\right)
-2\left(6M_{top}^2-G_U\right)+\frac{1}{2}G_U,\nonumber\\
&=&-G_N+\frac{5}{2}\left( 8\pi^2 \frac{d}{dt} \log B ^2\right)
-2\left(8\pi^2 \frac{d}{dt} \log M_{top}^2\right)+\frac{1}{2}G_U,\\
8\pi^2 \frac{d}{dt} \log M_{\nu_{\mu}}^2&=&-G_N+\frac{3}{2}\left( 8\pi^2 \frac{d}{dt}
 \log B ^2\right)-\left(8\pi^2 \frac{d}{dt} \log M_{top}^2\right)+\frac{1}{2}G_U,\\
8\pi^2 \frac{d}{dt} \log M_{\nu_{\tau}}^2&=&-G_N+\left( 8\pi^2 \frac{d}{dt}
 \log B ^2\right)-\frac{1}{2}\left(8\pi^2 \frac{d}{dt} \log M_{top}^2\right)+
\frac{1}{2}G_U. \label{e123}
\end{eqnarray}
The neutrino masses are obtained as 
\begin{eqnarray}
M_{{\nu}_{e}}=M_Ua_N^{-1/2}a_U^{1/4}{\left(\frac{B_Z}{B_U}\right)}^{5/2}
{\left(\frac{M_U}{M_{top}}\right)}^{2}\simeq{\lambda_{{\nu}_{e}}}^{16}\simeq
3\times10^{-9} \texttt{GeV},\\
M_{{\nu}_{\mu}}=M_Ua_N^{-1/2}a_U^{1/4}{\left(\frac{B_Z}{B_U}\right)}^{3/2}
{\left(\frac{M_U}{M_{top}}\right)}\simeq{\lambda_{{\nu}_{\mu}}}^{9}\simeq
1.4\times10^{-4} \texttt{GeV},\\
M_{{\nu}_{\tau}}=M_Ua_N^{-1/2}a_U^{1/4}{\left(\frac{B_Z}{B_U}\right)}
{\left(\frac{M_U}{M_{top}}\right)^{1/2}}\simeq{\lambda_{{\nu}_{\tau}}}^{6}\simeq
1.3\times10^{-2} \texttt{GeV}.\label{e129}
\end{eqnarray}
This completes our calculation of the masses of the quarks and leptons in terms of known 
and calculable quantities $\eta_F$, satisfying the RG Equations. 
It is absolutely necessary to write down the solutions of the RG
Equations for each quark and leptons separately, then alone, the 
contributions from the gauge interactions and the Yukawa mass coefficients $A_{FH}$
can be ascertained.

The values of the coefficients $\eta_F$ used above have been
calculated, upto to the nearest fraction, from the cited rotational integers $n_F$ 
in Table-\ref{tab:table3} and the top 
coupling coefficients $N_F$ from the RG Equations. They are given in the Table-\ref{tab:table4}.
\begin{center}
\begin{table}[h]
\caption{\label{tab:table4}Coefficients of $\eta_F$ for fermions}
\begin{tabular}{|c|c|c|c||c|c|c|c|}\hline\hline
quarks & $n_F$ & $N_F$ & $\eta_F$ &leptons & $n_F$ & $N_F$ & $\eta_F$\\ \hline
c&3&3&0&e&8&0&30/24\\ \hline
u&7&3&1/2&$\mu$&4&0&17/24\\ \hline
b&2&1&1/4&$\tau$&3&0&11/24\\ \hline
s&4&0&3/4&$\nu_{e}$&16&3&2\\ \hline
d&6&0&13/12&$\nu_{\mu}$&9&3&1\\ \hline
&&&&$\nu_{\tau}$&6&3&1/2\\ \hline
\end{tabular}
\end{table}
\end{center}

\section{Running masses of the fermions}
The exact one loop solution of the RGE for change in mass values with energy is given by
equation(\ref{eq24}). With $Z_F(t)=8\pi^2 \frac{d}{dt} \log B_F^2(t)$, this equation
reduces to
\begin{equation}
\frac{M_{top}^2}{M_F^2(t)}=\frac{M_{top}^2}{M_U^2}\frac{B_F^2(t_X)}{B_F^2(t)}
+\frac{A_F}{8\pi^2}\int_t^{t_X}dt_1 \frac{B_F^2(t_1)}{B_F^2(t)},\label{e130}
\end{equation} 
and
\begin{equation}
M_F(t)=M_F(M_X)\frac{B_F(t)/B_F(t_X)}{{\left(1+\frac{M_{U}^2}{M_{top}^2}
\frac{A_F}{8\pi^2}\int_t^{t_X}dt_1 \frac{B_F^2(t_1)}{B_F^2(t_X)} \right)}^{1/2}}.\label{eq130a}
\end{equation}
If we take
\begin{equation}
B_F(t)=\lambda^{n_F (1-t/t_X)}= \exp \left({n_F (1-t/t_X)}\log_e \lambda \right),
\end{equation}
and
\begin{equation}
M_F(t)=M_U\lambda^{n_F}{\left(1+\frac{M_{U}^2}{M_{top}^2}
\frac{A_F}{8\pi^2}\int_t^{t_X}dt_1\lambda^{2n_F (1-t_1/t_X)}\right)}^{1/2},
\end{equation}
with
\begin{eqnarray}
\int_t^{t_X}dt_1\lambda^{-2n_F}t_1/t_X&=&\int_t^{t_X}dt_1 e^{{-2n_F}t_1/t_X
 \log _e \lambda}\nonumber\\
&=&-\frac{t_X}{2n_F} \frac{1}{\log _e \lambda} \left[
e^{-2n_F \log _e \lambda} -e^{-2n_F{t/t_X} \log _e \lambda}\right]\nonumber\\
&=&-\frac{t_X}{2n_F}\frac{1}{\log _e \lambda}\left[ \lambda^{-2n_F}-
\lambda^{{-2n_F} t/t_X}\right],
\end{eqnarray}
we have, from equation (\ref{eq130a})
\begin{equation}
M_F(t)=M_F(M_X)\frac{\lambda^{n_F(1-t/t_X)}}{{\left[1+\frac{M_{F}^2(M_X)}{M_{top}^2}
\frac{A_F}{8\pi^2}\lambda^{2n_F}\lambda^{{2n_F}/t_X}
\left(\lambda^{-2n_F}-\lambda^{{-2n_F} t/t_X}\right) \right]}^{1/2}}\label{e130a}
\end{equation}

We shall not use this for constructing the tables. A simpler  approximate form 
ignoring $A_F$ terms of equation (\ref{e130a}),
\begin{equation}
M_F(t)=M_F(M_X) \lambda^{n_F \left(1-\frac{t}{33}\right)}
\end{equation} 
is presented in tabular form in Table-\ref{tab:table5} to Table-\ref{tab:table7}.
This gives the same CKM matrix.
Now, we present our results in the following tables.
\begin{center}
\begin{table}[h]
\caption{\label{tab:table5}Variation of $ M_{top}(t)$ with t= $log_e(\mu/M_Z)$}
\begin{tabular}{|c|c|c|c|c|c|c|c|c|c|c|c|c|}\hline
t       &0&3&6&9&12&15&18&21&24&27&30&33\\ \hline
$M_{top}(t)$(GeV)&175.88&166.98&159.60&153.25&147.61&142.46&137.64&133.02&128.
51&124.51&119.56&115.00\\\hline
\end{tabular}
\end{table}
\end{center}
From the above values, we find that $\frac{d^2M_{top}(t)}{dt^2}$ is nearly zero. The
top mass variation is approximately
\begin{equation}
M_{top}(t)=\left [ 175.-\frac{t}{33}\times 60\right ] ~~\texttt{GeV}.
\end{equation}

\begin{figure}[b]
\begin{center}
\includegraphics[width=11.0cm,height=8.0cm]{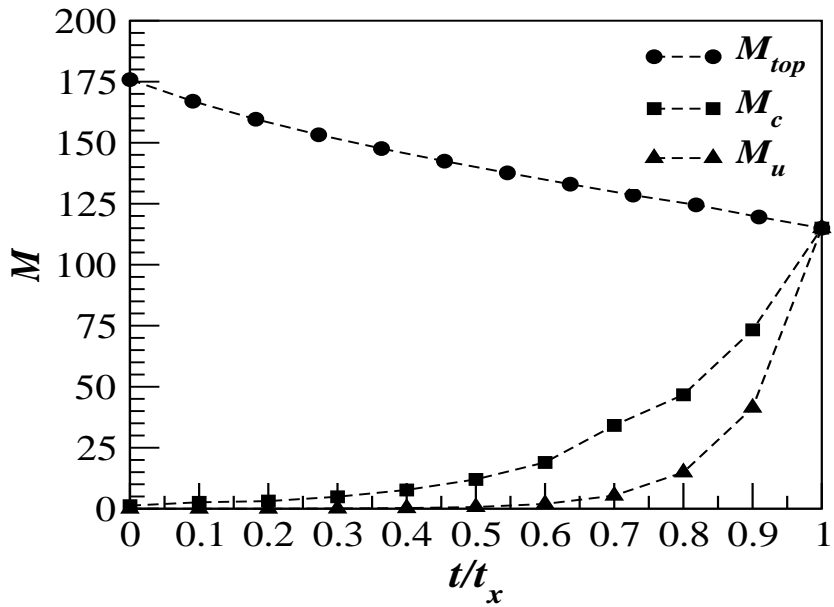}
\caption{\label{fig:Fig1}Variation of masses of Up-quarks in Gev with t/$t_X$ ,
t=log($\mu/M_Z$) and $t_X$=33.} 
\end{center}
\end{figure}

\begin{figure}[h]
\begin{center}
\includegraphics[width=9.6cm,height=7.6cm]{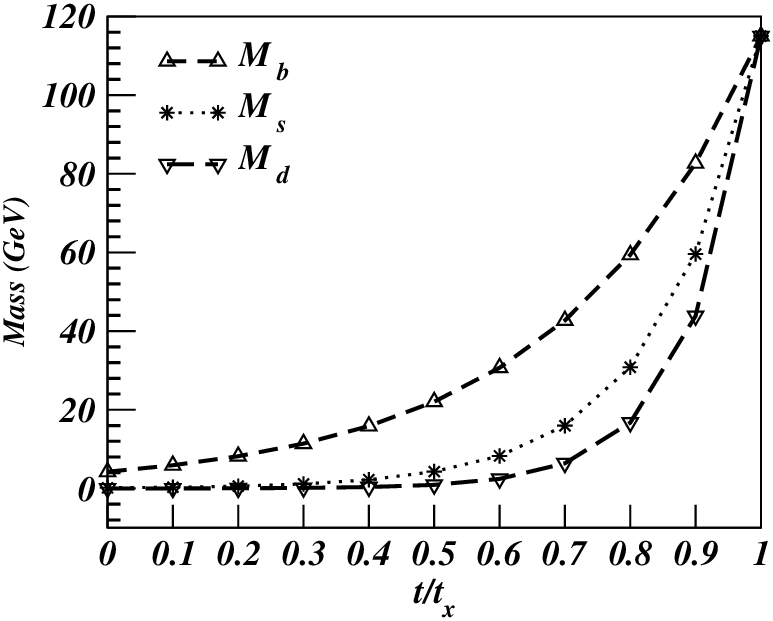}
\caption{\label{fig:Fig2}Variation of masses of Down-quarks in Gev with 
t/$t_X$ , t=log($\mu/M_Z$) and $t_X$=33.} 
\end{center}
\end{figure}

\begin{figure}[b]
\begin{center}
\includegraphics[width=9.5cm,height=6.8cm]{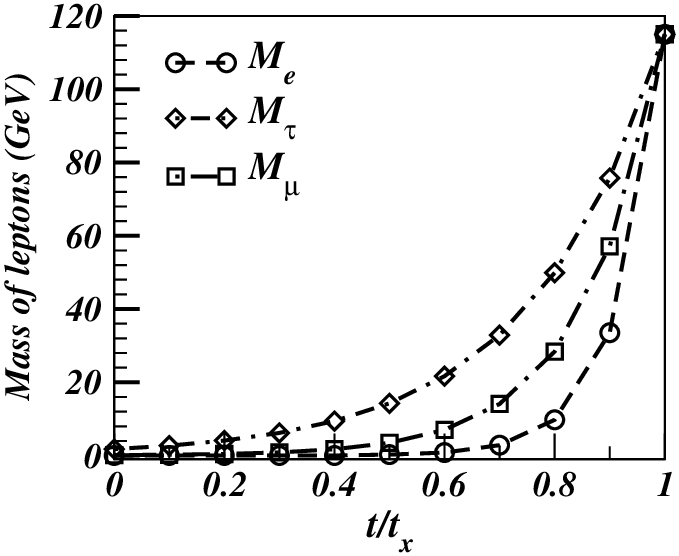}
\caption{\label{fig:Fig3} Variation of masses of Electrons in Gev with 
t/$t_X$ , t=log($\mu/M_Z$) and $t_X$=33.} 
\end{center}
\end{figure}

\begin{figure}[h]
\begin{center}
\includegraphics[width=10.0cm,height=7.8cm]{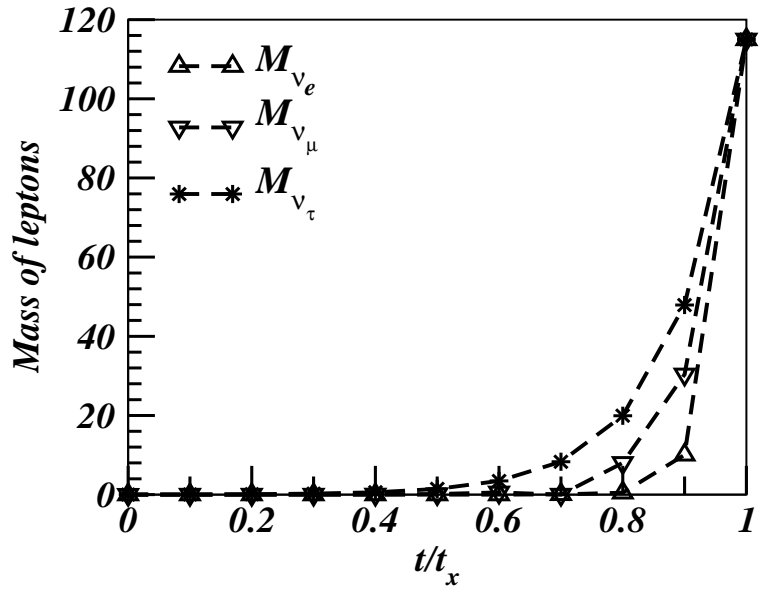}
\caption{\label{fig:Fig4}Variation of masses of Neutrinos in Gev with 
t/$t_X$ , t=log($\mu/M_Z$) and $t_X$=33.} 
\end{center}
\end{figure}

\begin{figure}[b]
\begin{center}
\includegraphics[width=16.1cm,height=10.5cm]{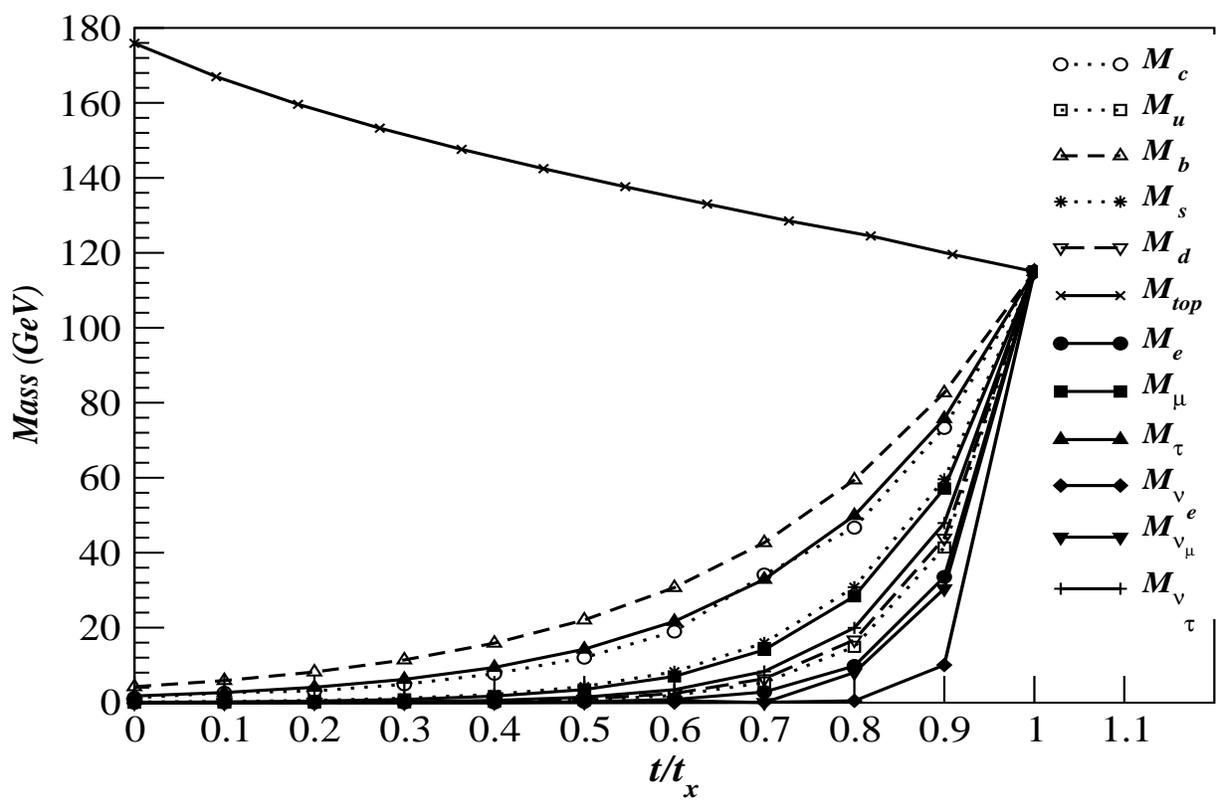}
\caption{\label{fig:Fig5}Variation of masses of all 12 fermions in Gev with 
t/$t_X$ , t=log($\mu/M_Z$) and $t_X$=33. } 
\end{center}
\end{figure}

\begin{center}
\begin{table}[h]
\caption{\label{tab:table6}Running Lepton Masses in GeV, $t_X$=33. }
\begin{ruledtabular}
\begin{tabular}{|l|l|l|l|l|l|l|}
t/$t_X$ & $M_{e}$ & $M_{\mu}$ &$M_{\tau}$& $M_{\nu_{e}}$&$M_{\nu_{\mu}}$
& $M_{\nu_{\tau}}$\\
\hline
0& 0.00051 & 0.105 & 1.77 & 3$\cdot 10^{-9}$ & 1.9$\cdot10^{-4}$ & 0.018\\
0.1& 0.00175 & 0.2115 & 2.687 & 3.43$\cdot 10^{-8}$&7.19$\cdot 10^{-4}$ & 0.043\\
0.2& 0.006 & 0.426 & 4.079 & 3.92$\cdot 10^{-7}$ & 2.72$\cdot 10^{-3}$ & 0.104\\
0.3& 0.0206 & 0.8575 & 6.192 & 4.49$\cdot 10^{-6}$ & 1.03$\cdot 10^{-2}$ & 0.25\\
0.4& 0.02 & 1.726& 9.399 & 5.13$\cdot 10^{-5}$ & 3.9$\cdot 10^{-2}$  & 0.6\\
0.5& 0.242 & 3.476 & 14.267 &5.87$\cdot 10^{-4}$ & 0.1478 & 1.44\\
0.6& 0.830 & 6.999 & 21.658 & 6.72$\cdot 10^{-3}$  &  0.56 &3.46\\
0.7& 2.849 & 14.081 & 32.876 &  7.68$\cdot 10^{-2} $ & 0.01 &8.3\\
0.8& 9.774 & 28.369 &49.910 & 0.482& 8.02 & 19.93\\
0.9& 33.527 & 57.118& 75.758& 10.05 & 30.374 & 47.9\\
1.0& 115.0& 115.0 & 115.0 &115.0 &115.0 &115.0\\
\end{tabular}
\end{ruledtabular}
\end{table}
\end{center}
\begin{center}
\begin{table}[h]
\caption{\label{tab:table7}Values of Running quark masses in GeV
(other than the top quark), $t_X$=33.}
\begin{ruledtabular}
\begin{tabular}{|l|l|l|l|l|l|}
t/$t_X$ & $M_{c}$ & $M_{u}$ &$M_{b}$& $M_{s}$&$M_{d}$\\
\hline
0.0& 1.27 & 0.0042 & 4.23 & 0.159& 0.0075 \\
0.1& 2.62 & 0.0116 & 5.88 & 0.307&0.0196 \\
0.2& 3.12 & 0.0324 & 8.18 & 0.593 & 0.051\\
0.3& 4.9 &0.09  &11.39 &1.146  & 0.135 \\
0.4& 7.7 &0.25 &15.85 &2.213 &0.354 \\
0.5& 12.0&0.695&22.05&4.276&0.9287 \\
0.6& 18.96&1.9307&30.687&8.26&2.434\\
0.7& 34.16&5.3636&42.696&15.95&6.383\\
0.8& 46.69&14.9&59.406&30.82&16.73\\
0.9& 73.28&41.395&82.651&59.59&43.86\\
1.0& 115.0& 115.0 & 115.0 &115.0 &115.0\\
\end{tabular}
\end{ruledtabular}
\end{table}
\end{center}
The graphs are much more revealing.
The variation of mass for the Up quarks, including the top,
have been shown in figure-\ref{fig:Fig1}, for the Down in figure-\ref{fig:Fig2}
, for the Electrons in figure-\ref{fig:Fig3} and for the Neutrinoes in figure-
\ref{fig:Fig4}. All have been compressed in figure-\ref{fig:Fig5} to allow a 
glance at the totality of the descent and ascent of the masses 
to the unification mass of 115 GeV. We believe that an exact analysis will not 
differ much from those presented in these figures.

\section{Concluding Remarks}
Deo and Maharana\cite{a10} have already given exact solutions for the 
one loop RGEs in MSSM. They proved that all the 
fermions might have originated from a common mass $M_U\simeq$ 115 GeV at the GUT energy of 
$M_X\simeq 2.2\times 10^{16}$ GeV in a perturbative scheme. This has not been confirmed
by authors working in this field as they have not looked at the solutions which 
proved the same wayas one loop gauge sector renormalisation group equation.

As the energy diminishes, the mass of the top increases to ~ 175 GeV at
the $Z$-meson mass $M_Z$, whereas, masses of all other quarks and tiny
leptonic masses increase from their value at $M_Z$ to $M_U$ of each 115 GeV.
Hopefully, the exact nature of this variation could be obtained from RGE. By introducing
an auxiliary function $B(t)$ through the mixing terms of RGE,
a very simplified expression can be obtained for both the mass values at
$M_Z(t=0)$ and their variation till $M_X$ when, the masses of the fermions
which keep changing, attain the final value at $M_X$.

The success of the analysis of the Wolfenstein and the ratio parametrization
of RRR given by equation(\ref{eq8}) is clearly brought out by our solutions. 
An extension to running masses at linearity level, leads to a very simple formula for 
the fermions other than the top, 
\begin{equation}
M_F(t)=M_U \left (\frac{M_F}{M_U}\right )^{(1-\frac{t}{t_X})}=
M_U~ \lambda^{n_F(1-\frac{t}{t_X})},
\end{equation}
where 
$M_F^{exp.}=M_F(M_Z)=\lambda^{n_F}$. The increase is exponential. 
Much more exact analytic studies are needed to
deduce  the values of the $n_F$ from RGE. As a first step, we follow up
the texture analysis procedure by introducing a similar, but not the same
function $B(t)$\cite{a8}, which is such that $d \log B(t)=d \log M_{top}(t)$.
The resulting non-uniqueness is taken advantage of. Introducing known gauge
couplings and the values of $n_F$, the effect of gauge interaction
have been calculated.

In the foregoing sections 1 to 8, there has been two important omissions.
We have not remarked neither about higher loop effects nor about 
the threshhold corrections. They may be very large due to very heavy top.
There has been no mention of the CKM phases. But, we work with three generations
only. So, in this case there is precisely, one phase angle $\phi$. This is
defined from the Wolfenstein parametrisation as
\begin{equation}
\lambda= \left ( \frac{M_d}{M_s} +\frac{M_u}{M_c} +
2\sqrt{\frac{M_d}{M_s}\frac{M_u}{M_c}}\cos\phi\right )^{\frac{1}{2}},
\end{equation}
and can be deduced by extending the Gatto-Sarton-Tonio-Oakes
(GSTO)\cite{a14} to next leading order of solution of RGE, $\lambda =
(M_d/M_s)^{1/2}$. Putting in the numbers $n_F$ from table-III,  we get
\begin{equation}
\cos\phi\simeq -\frac{\lambda}{2}\simeq -0.1,~~~~~~\phi\simeq 95^0.
\end{equation}
This result is reasonable.

It may be argued that there are too many parameters still in the guise of 
the rotation integers $n_F$ even though $\lambda$ could be determined. Let us
recall the case of the unification of the strong, the weak, the electromagnetic
 and the gravitational interactions. One consciously omits gravity and lists 
the other three in order of their strengths. Here the  top mass, with 
the largest value,has been deduced from the knowledge of the unification mass
 and gauge couplings quite accurately. For $F$=2, the two successive rotations
i.e., 0 to $4\pi$ gives $n_F$=2 and $M_U {\lambda}^2$ falls near the mass of the
bottom. This process of complete successive rotations yields the masses of all
the quarks and electrons, except $n_F$=5. However, there is a GUT prediction
by Georgi and Jarlskog\cite{a3}, that the ratio $M_s/M_d\simeq 25\simeq{\lambda}^{-2}$.
This may be the reason why $n_F$ jumps from 4 to 6. There is also another accurate
prediction
\begin{equation}
\frac{M_d/M_s}{(1-M_d/M_s)^2} =9.\frac{M_e/M_{\mu}}{(1-M_e/M_{\mu})^2}.
\end{equation}
With these GUT supplements, there is no other unknown additional parameter left.
The unification mass $M_U\simeq 115$ GeV is the only input. 

Corroborating and in consonance with a lot of excellent work in neutrino 
physics have been done by many workers ,e.g. Babu, Mohapatra and Barr\cite{a15}. 
The neutrino masses  fall with the rotational integers, determined for the 
neutrinoes in the table-V in a very nice way.

The figure-5  illustrates the significance of this approach of solving the RGE.
The similarity of figure-5 with gauge unification graph is striking. Thus, we prove,
beyond doubt that the original ideas of Pati and Salam\cite{a1} is correct as 
well as the SUSY standard model. We have given many extrapolatory ideas
which can be successfully implemented so that SUSY standard model world can be
characterised by only three parameters, the GUT mass $M_X\simeq 2.2\times 10^{16}$,
 the GUT coupling strength $\alpha_{GUT} \simeq \frac{1}{24.5}$ and common-origin
-fermion mass $M_U \simeq115$ GeV. This is a simplification beyond expectation and
very much gratifying.

\begin{acknowledgments}
We thank Prof. L. Maharana and Dr. P.K. Jena for elaborate discussions and Sri
Sidhartha Mohanty and Sri Haraprasanna Lenka for help in computations  and 
in preparing graphs respectively.
\end{acknowledgments}

\end{document}